\documentclass[11pt]{article}

\PassOptionsToPackage{usenames,dvipsnames}{color}
\usepackage{fullpage}
\usepackage[T1]{fontenc}
\usepackage[usenames,dvipsnames]{xcolor}
\usepackage{stmaryrd}

\usepackage{mathpazo}
\usepackage{xfrac}
\usepackage{bm}
\usepackage{todonotes}
\usepackage{setspace}
\usepackage{scrextend}
\usepackage{enumitem}

\usepackage{atmacros}

\RequirePackage[colorlinks=true,backref=page]{hyperref}
\hypersetup{
  linkcolor=[rgb]{0.3,0.3,0.6},
  citecolor=[rgb]{0.2, 0.6, 0.2},
  urlcolor=[rgb]{0.6, 0.2, 0.2}
}

\usepackage{amsthm}
\usepackage{thmtools,thm-restate}

\numberwithin{equation}{section}
\declaretheoremstyle[bodyfont=\it,qed=\qedsymbol]{noproofstyle}

\declaretheorem[numberlike=equation]{observation}

\declaretheorem[name=Observation,numbered=no]{observation*}

\declaretheorem[numberlike=equation]{theorem}

\declaretheorem[name=Theorem,numbered=no]{theorem*}

\declaretheorem[numberlike=equation]{lemma}
\declaretheorem[name=Lemma,numbered=no]{lemma*}

\declaretheorem[numberlike=equation]{corollary}
\declaretheorem[name=Corollary,numbered=no]{corollary*}

\declaretheorem[name=Proposition,numbered=no]{proposition*}

\declaretheorem[name=Claim,numbered=no]{claim*}

\declaretheorem[name=Conjecture,numbered=no]{conjecture*}

\declaretheorem[numberlike=equation]{question}
\declaretheorem[name=Question,numbered=no]{question*}

\declaretheoremstyle[bodyfont=\it,qed=$\lozenge$]{defstyle} 

\declaretheorem[numberlike=equation,style=defstyle]{definition}
\declaretheorem[unnumbered,name=Definition,style=defstyle]{definition*}

\declaretheorem[unnumbered,name=Example,style=defstyle]{example*}

\declaretheorem[unnumbered,name=Notation=defstyle]{notation*}

\declaretheorem[unnumbered,name=Construction,style=defstyle]{construction*}

\declaretheorem[numberlike=equation,style=defstyle]{remark}
\declaretheorem[unnumbered,name=Remark,style=defstyle]{remark*}

\newcommand{\ignore}[1]{}

\newcommand{\inparen}[1]{\left( #1 \right)}
\newcommand{\insquare}[1]{\left[ #1 \right]}
\newcommand{\inbrace}[1]{\left\{ #1 \right\}}

\newcommand{\inangle}[1]{\left< #1 \right>}

\newcommand{\set}[1]{\inbrace{#1}}

\newcommand{\abs}[1]{\left| #1 \right|}

\newcommand{\coeff}{\operatorname{coeff}}

\newcommand{\cvector}{\overline{\operatorname{coeff}}}

\newcommand{\C}{\mathbb{C}}

\newcommand{\Det}{\operatorname{\sf Det}}
\newcommand{\Perm}{\operatorname{\sf Perm}}

\newcommand{\commRO}[1]{\mathrm{commRO}\inparen{#1}}
\newcommand{\diagRO}[1]{\mathrm{diagRO}\inparen{#1}}
\newcommand{\DPD}[1]{\dim \partial^{<\infty}\inparen{#1}}

\newcommand{\V}{\mathbf{V}}
\newcommand{\I}{\mathbf{I}}

\newcommand{\normalset}{\operatorname{N}}

\newcommand{\trans}[1]{{#1}^{\intercal}}

\newcommand{\LM}{\operatorname{LM}}

\newcommand{\vecA}{\mathbf{A}} 
\usepackage{nth}
\usepackage{intcalc}
\usepackage{etoolbox}
\usepackage{xstring}

\usepackage{ifpdf}
\ifpdf
\else
\usepackage[quadpoints=false]{hypdvips}
\fi

\newcommand{\ECCCurl}[2]{http://eccc.hpi-web.de/report/\ifnumcomp{#1}{>}{93}{19}{20}#1/#2/}

\newcommand{\shortECCC}[2]{\texttt{\href{http://eccc.hpi-web.de/report/\ifnumcomp{#1}{>}{93}{19}{20}#1/#2/}{eccc:TR#1-#2}}}

\newcommand{\parseECCC}[1]{
\StrSubstitute{#1}{TR}{}[\tmpstring]%
\IfSubStr{\tmpstring}{/}{ 
\StrBefore{\tmpstring}{/}[\ecccyear]%
\StrBehind{\tmpstring}{/}[\ecccreport]%
}{
\StrBefore{\tmpstring}{-}[\ecccyear]%
\StrBehind{\tmpstring}{-}[\ecccreport]%
}%
\shortECCC{\ecccyear}{\ecccreport}}

\title{Explicit Commutative ROABPs from Partial Derivatives}

\author{
 	Vishwas Bhargava \thanks{David R. Cheriton School of Computer Science, Univeristy of Waterloo, Waterloo, Canada. Email: \texttt{vishwas1384@gmail.com}.}
 	\and
 	Anamay Tengse \thanks{School of Computer Sciences, NISER, Bhubaneswar. A major part of this work was done as a postdoc at the Univeristy of Haifa (ISF grant no. 716/20), and Reichman University, Herzliya (ISF grant no. 843/23). Email: \texttt{anamay.tengse@gmail.com}.}
}

\date{\today}

\onehalfspace

\begin{document}

\maketitle


\begin{abstract}
    The dimension of partial derivatives (Nisan and Wigderson, 1997) is a popular measure for proving lower bounds in algebraic complexity.
    It is used to give strong lower bounds on the \emph{Waring decomposition} of polynomials (called \emph{Waring rank}).
    This naturally leads to an interesting open question: does this measure essentially characterize the Waring rank of any polynomial?
    
    The well-studied model of Read-once Oblivious ABPs (ROABPs for short) lends itself to an interesting hierarchy of `sub-models': Any-Order-ROABPs (ARO), Commutative ROABPs, and Diagonal ROABPs.    
    It follows from previous works that for any polynomial, a bound on its Waring rank implies an analogous bound on its Diagonal ROABP complexity (called the \emph{duality trick}), and a bound on its dimension of partial derivatives implies an analogous bound on its `ARO complexity': ROABP complexity in any order (Nisan, 1991).
    Our work strengthens the latter connection by showing that a bound on the dimension of partial derivatives in fact implies a bound on the commutative ROABP complexity.
    Thus, we improve our understanding of partial derivatives and move a step closer towards answering the above question.

    Our proof builds on the work of Ramya and Tengse (2022) to show that the \emph{commutative-ROABP-width} of any homogeneous polynomial is at most the dimension of its partial derivatives.
    The technique itself is a generalization of the proof of the \emph{duality trick} due to Saxena (2008).
\end{abstract}

\section{Introduction}

How many points do we need to evaluate an expression like the following on, to deterministically tell if it is computing the zero polynomial?
\begin{equation}\label{eq:typical-waring}
  f(x_1,\ldots,x_n) = \inparen{a_{1,1} x_1 + a_{1,2} x_2 + \cdots + a_{1,n} x_n}^d + \cdots + \inparen{a_{s,1} x_1 + \cdots + a_{s,n} x_n}^d
\end{equation}

As of today, the answer to this question stands at $(nds)^{O(\log\log n)}$, just a (rather annoying) smidgen away from a legit efficient algorithm.
The above bound follows from a combination of the works of Forbes, Saptharishi and Shpilka~\cite{FSS14} and Gurjar, Korwar and Saxena~\cite{GKS17}.

An expression like \eqref{eq:typical-waring} is called a \emph{Waring decomposition for $f$} of size $s$; the name comes from `Waring's problem' in number theory\footnote{See \href{https://en.wikipedia.org/wiki/Waring\%27s_problem}{this \texttt{wikipedia} article} for a summary.}.
Analogously, for a homogeneous polynomial $f(\vecx)$ of degree $d$, its \emph{Waring rank} is the smallest number $s$ for which $f$ can be written as a sum of $d$-th powers of $s$-many linear forms; that is, the size of its smallest Waring decomposition.
The Waring rank of different polynomials has been studied in mathematics for over a century now (see e.g. \cite{IK99}), and some recent works have even found its applications in parameterized algorithms (e.g. \cite{P19}).
It is known that any polynomial has a finite Waring rank \cite{F94,CCG12}, except for over finite fields of small characteristic.
We now also know the Waring rank of monomials exactly~\cite{RS11}.
For example, it is known that the monomial $x_1 x_2 \cdots x_n$ has Waring rank exactly $2^{n-1}$.

The corresponding algebraic model of computation: called a ``depth-3-powering circuit'', was first introduced in algebraic circuit complexity by Saxena~\cite{S08b}, who studied it from the perspective of polynomial identity testing (PIT for short).
PIT is the algorithmic task mentioned above: determine whether the given circuit computes the identically zero polynomial.
In what is sometimes called a ``whitebox PIT'', the algorithm has access to the circuit itself; Saxena~\cite{S08b} gave an efficient whitebox test for a more general model.
In a ``blackbox PIT'', the algorithm cannot access the expression and can only query it on a few points (independent of the actual circuit), which is exactly the question stated at the start.

\subsubsection*{Dimension of partial derivatives}

All the currently known blackbox PITs for depth 3 powering circuits build on the fact that any $n$-variate, degree-$d$ polynomial with Waring rank $s$ has at most $s(d+1)$ \emph{dimension of partial derivatives} (see \autoref{defn:dimension-of-partials}).
The measure was introduced by Nisan and Wigderson~\cite{NW97} as a tool to prove lower bounds against sums of \emph{products} of linear forms, and thus the above statement is implicit from their work.
The myriad variants of this measure now form the basis of several strong lower bounds throughout algebraic circuit complexity (see e.g. \cite{SY10,S15}).

Returning to Waring decompositions, almost all known lower bounds on Waring ranks of different polynomials use the dimension of partial derivatives in one way or the other.
In view of this, and given the strong connections between proofs of hardness and derandomization of PIT (see e.g. \cite{KS19}), it stands to reason that obtaining an efficient blackbox PIT for depth 3 powering circuits requires us to answer the following question.
\begin{question}\label{quest:DPD-captures-Waring}
  Is it the case that any $n$-variate polynomial with dimension of partial derivatives $r$ has a Waring rank that is at most $\poly(n,r)$?
\end{question}
To the best of our knowledge, there aren't even any candidate negative examples to this question, except for the symbolic determinant: $\Det_n$.
The $n \times n$ determinant has a dimension of partial derivatives that is $2^{\Theta(n)}$, but the best known upper bound on its Waring rank stands at $2^{O(n \log n)}$.

The only other ``deviation'' that these two measures --- dimension of partial derivatives and Waring rank --- exhibit, comes from their respective connections with a different well-studied model, which we will now see.

\subsection{Read-once Oblivious ABPs (ROABPs)}

An ROABP is an expression of the form: $\trans{\vecu} \cdot M_1(x_1) \cdot M_2(x_2) \cdots M_n(x_n) \cdot \vecv$, where $\vecu, \vecv$ are vectors over the base field and each $M_j(x_j)$ is a univariate polynomial with matrices as coefficients, as follows.
\begin{equation}\label{eq:typical-roabp-layer-matrix}
  M_j(x_j) = A_{j,0} + A_{j,1} x_i + A_{j,2} x_i^2 + \cdots + A_{j,d} x_j^d
\end{equation}
That is, there is exactly one `matrix-polynomial' corresponding to each variable.
Thus, each variable is ``read'' exactly once, oblivious to the other variables; hence the name.
We formally define ROABPs in \autoref{defn:ROABP}.

Here, the dimension of $\vecu$, $\vecv$ and all the $n(d+1)$ many matrices (assumed to be the same without loss of generality) is said to be the \emph{width} of the ROABP, and is typically denoted by $w$.
Note that the width of an ROABP is the single parameter that dictates its complexity, since $n$ and $d$ arise straight from the polynomial being computed.
A subtle point here is that for the same polynomial, the smallest possible ROABP-width can vary widely depending on the order in which the variables appear (see \autoref{obs:pallindrome-order-dependence}), and hence the order of an ROABP is also an important parameter.
Nevertheless, for any polynomial and any order, the exact size of the smallest corresponding ROABP can be obtained using a characterization given by Nisan~\cite{N91}.
As we will soon see, even this characterization is in a way connected to the partial derivatives of the given polynomial; we provide a formal definition and statement in \autoref{defn:Nisan-matrix} and \autoref{prop:Nisan-characterization}.

ROABPs were formally introduced by Forbes and Shpilka~\cite{FS13} as algebraic analogues of ROBPs from the boolean world, where they showed a quasi-polynomial time blackbox PIT for ROABPs, inspired by Nisan's PRG construction against ROBPs~\cite{N92}.
As the name suggests, ROABPs are a special case of ``algebraic branching programs'' which are an algebraic analogue of the (boolean) branching programs.
However, we omit those definitions of ABPs and ROABPs, as seeing them as the matrix-vector product expressions like above would be more useful for the discussions in this paper.
We now introduce the structured variants of ROABPs that are relevant to this work.

\subsection{Variants of ROABPs}

Since the ROABP-complexity of some polynomials depends heavily on the underlying order, we can further cut out a subclass of polynomials that admit $\poly(n,d)$-sized ROABPs: those that admit $\poly(n,d)$-sized ROABPs \emph{in every order}.
This class of polynomials is sometimes referred to as ``Any-order ROABPs'' (AROs for short)\footnote{Contrary to what the name suggests, this is not a special type of ROABPs, it is a class of polynomials.}.

A \emph{syntactic} way of ensuring that a polynomial computed by an efficient ROABP belongs to ARO, is to ensure that all the $n(d+1)$-many coefficient matrices ($A_{j,\ast}$s in \eqref{eq:typical-roabp-layer-matrix}) commute with each other under multiplication.
This then means that for any $j,j' \in [n]$, we have that $M_j(x_j) M_{j'}(x_{j'}) = M_{j'}(x_{j'}) M_j(x_j)$, and then the layers of the same ROABP can be shuffled to work for any order.
Such an ROABP with commuting coefficient matrices is called a \emph{commutative ROABP} (commRO for short); a formal definition is in \autoref{defn:commRO}.

Finally, an easy way to pick coefficient matrices that commute with each other is to choose all of them as diagonal matrices.
Such an ROABP is called a \emph{diagonal ROABP} (diagRO for short), defined in \autoref{defn:diagRO}.

The above variants of ROABPs appear implicitly in some previous works on ROABPs, but they were explicitly defined and proposed as objects of study in a recent work of Ramya and Tengse~\cite{RT22}.
As mentioned earlier, our proof technique also borrows from the algebraic machinery that appears in their work.

We now proceed to look at the connections between Waring rank, dimension of partial derivatives and these structured ROABPs, before stating our main result.

\subsection{Waring rank, partial derivatives and ROABPs}

The aforementioned whitebox PIT for depth 3 powering circuits due to Saxena~\cite{S08b} has the following result at its core.
\begin{theorem}[{Duality trick~\cite[Lemma 1]{S08b} (Informal)}]\label{prop:duality-trick-informal}
  For any linear form $\ell(\vecx) = a_1 x_1 + a_2 x_2 + \cdots + a_n x_n$, and any $d$, the polynomial $\ell(\vecx)^d$ can be expressed as:
  \[
      \ell(\vecx)^d = \sum_{i = 1}^t \beta_i \cdot g_{i,1}(x_1) \cdot g_{i,2}(x_2) \cdots g_{i,n}(x_n),
  \]
  for constants $\beta_1,\ldots,\beta_t$ and degree-$d$ univariates $g_{1,1},\ldots,g_{t,n}$, with $t \leq nd + 1$.
\end{theorem}

Note that this gives an ROABP of width $t = O(nd)$ for the $d$-th power of any $n$-variate linear form, by using the $g_{i,j}$s appropriately to obtain each of the matrix polynomials $M_j(x_j)$ and using $\beta_i$s in the vector $\vecu$ (or $\vecv$).
In fact, the ``coefficient matrices'' (the $A_{j,\ast}$s from \eqref{eq:typical-roabp-layer-matrix}) of this ROABP are just diagonal matrices.
Consequently, an $n$-variate, degree-$d$ polynomial with Waring rank $r$ has a diagRO of width $O(ndr)$.

The duality trick actually provides diagROs for a more general model called ``depth 4 diagonal circuits'', and the corresponding whitebox PIT also holds for this more general model.
In fact, this relation between powering circuits and diagROs is a crucial component of the current state-of-the-art blackbox PIT for depth 3 powering circuits~\cite{FSS14,GKS17} mentioned earlier.

Given that polynomials with small Waring rank have small diagROs, it is natural to ask what happens to polynomials with small dimension of partial derivatives.

\paragraph*{ROABPs and partial derivatives.}
Suppose we are given an $n$-variate, degree-$d$ polynomial $f(\vecx)$, whose dimension of partial derivatives is at most $r$.
It turns out, via Nisan's characterization, that such a polynomial has an ROABP of width at most $r$ \emph{in every order} (see \autoref{obs:partials-and-Nisan}).
That is, for any $\sigma \in s_n$, we are guaranteed \emph{some} ROABP, say $R_{\sigma}(\vecx)$, that computes $f$ in that order.
However, it is not clear from this non-constructive upper bound whether the ROABPs $R_{\sigma}$ across different $\sigma$s are related in any way.
This brings us to our main result.

\subsection{Our contribution}

We first formally define the measure: dimension of partial derivatives.
\begin{definition}[Dimension of partial derivatives]\label{defn:dimension-of-partials}
  For a polynomial $f(x_1,\ldots,x_n)$, the dimension of its partial derivatives is defined as follows.
  \[
      \DPD{f} := \dim\inparen{ \mathrm{span}_{\C} \set{ \partial_{\vece} f : \vece \in \N^n } }
  \]
  Here $\partial_{\vece} f$ denotes the partial derivative of $f$ with respect to the monomial $\vecx^{\vece} = x_1^{e_1}x_2^{e_2}\cdots x_n^{e_n}$.
\end{definition}

For a polynomial $f$, let $\commRO{f}$ denote the width of the smallest commRO that computes it; our main result is as follows.
\begin{restatable}{theorem}{MainTheorem}\label{thm:main}
  For any homogeneous polynomial $f(\vecx) \in \C[\vecx]$, $\commRO{f} \leq \DPD{f}$.
\end{restatable}

Since the dimension of partial derivatives of any homogeneous component of $f$ is at most $\deg(f)$ times that of $f$ (see \autoref{lem:partials-homogeneous-component}), we get the following result in the general case.
\begin{corollary}\label{cor:general-polys}
  For any $f(\vecx) \in \C[\vecx]$ of degree $d$, $\commRO{f} \leq (d+1)^2 \cdot \DPD{f}$.
\end{corollary}

\paragraph*{Set multilinear upper bounds}

In fact, our method for constructing commutative ROABPs using $\DPD{f}$ also lets us obtain what we call ``commutative set-multilinear ABPs'' for $f$ with a minor tweak in our proof.
Informally, a degree-$d$ polynomial $f(\vecx)$ is called set-multilinear under a partition $\vecx = \vecx_1 \sqcup \vecx_2 \sqcup \cdots \sqcup \vecx_d$, if each of its monomials contains exactly one variable from each of the $\vecx_i$s.
A(n ordered) set-multilinear ABP is then a product of matrices with linear polynomials as entries, with the variables in those polynomials obeying the partition (see definitions \ref{defn:smABP} and \ref{defn:comm-smABP}).

\begin{restatable}{theorem}{commSMABP}\label{thm:comm-smABP}
    For any set-multilinear polynomial $f(\vecx) \in \mathbb{C}[\vecx]$, the commutative-set-multilinear-ABP-width$(f) \leq \DPD{f}$.    
\end{restatable}

\paragraph*{Explicitness.} We note that our proof provides an explicit construction of a commRO for any polynomial $f$, given the \emph{dependencies} between the partial derivatives of $f$.
In fact, as mentioned earlier, this construction itself is a generalization of the proof of the duality trick from Saxena's work~\cite{S08b} (see \autoref{rmk:connection-to-duality}).
We describe a width-$2^{O(n)}$ commRO, and a commutative set-multilinear ABP for the $n \times n$ determinant to illustrate this point in \autoref{sec:commRO-for-determinant}.

\section{Preliminaries}

Throughout the paper, we work with the field of complex numbers, but most of our proofs extend to fields whose characteristic is zero or large enough. 

\paragraph*{Notation}
\begin{itemize}
  \item For a vector $\vece \in \N^n$, we write $\vect^{\vece}$ for the monomial $t_1^{e_1} t_2^{e_2} \cdots t_n^{e_n}$, where $\vect$ is a set of variables.
  We also use $\vece !$ to refer to the product of factorials $e_1! e_2! \cdot e_n!$.
  \item For a monomial $m$, we write $\partial_m f$ for the partial derivative $\frac{\partial^{\abs{\vece}} f}{\partial \vect^{\vece}}$.
  When $m = \vect^{\vece}$, we shorten it further to $\partial_{\vece}f$.
\end{itemize}

\subsection{Formal definitions}

\begin{definition}[Read-once Oblivious ABP (ROABP)]\label{defn:ROABP}
  For any $n,d,w \in \N$, and an $n$-variate polynomial $f(\vecx)$ of individual degree $d$, we say that it has a width $w$ ROABP, if there exists a permutation $\sigma \in s_n$ for which there exist matrices $\set{A_{j,k}}$ in $\C^{w \times w}$ for all $j \in [n]$ and $0 \leq k \leq d$, and vectors $\vecu, \vecv \in \C^w$, such that the following holds.
  \begin{align*}
      f(\vecx) &= \trans{\vecu} \cdot M_{\sigma(1)}(x_{\sigma(1)}) \cdot M_{\sigma(2)}(x_{\sigma(2)}) \cdots M_{\sigma(n)}(x_{\sigma(n)}) \cdot \vecv,\\
      &\text{where for all $j \in [n]$,}\\
      M_{j}(x_{j}) &= A_{j,0} + A_{j,1} x_j + A_{j,2} x_j^2 + \cdots + A_{j,d} x_j^d.
  \end{align*}
  We call the matrices $\set{A_{j,k}}$ the \emph{coefficient matrices} of the ROABP.
\end{definition}

\begin{definition}[Commutative ROABP (commRO)]\label{defn:commRO}
  An ROABP is said to be a \emph{commutative ROABP}, if all its coefficient matrices commute with each other pairwise.

  For a polynomial $f$, we use $\commRO{f}$ to denote the smallest width $w$ such that there is width-$w$ commRO computing $f$.
\end{definition}

\begin{definition}[Diagonal ROABP (diagRO)]\label{defn:diagRO}
  An ROABP is said to be a \emph{diagonal ROABP}, if all its coefficient matrices are diagonal matrices.

  For a polynomial $f$, we use $\diagRO{f}$ to refer to the smallest width $w$ such that there is width-$w$ diagRO computing $f$.
\end{definition}
\paragraph*{Partial Derivatives and the Nisan matrix}

\begin{lemma}\label{lem:partials-homogeneous-component}
  Let $f(\vecx)$ be a polynomial of degree $d$ and let $h(\vecx)$ be some homogeneous component of $f$.
  Then $\DPD{h} \leq (d+1) \cdot \DPD{f}$.
\end{lemma}
\begin{proof}
  Note that for any nonzero scalar $\alpha$, the polynomial $f_{\alpha}(\vecx) := f(\alpha x_1, \alpha x_2, \ldots, \alpha x_n)$ satisfies $\DPD{f_{\alpha}} = \DPD{f}$, since it is an invertible operation.
  Next, for distinct $\alpha_0, \alpha_1, \ldots, \alpha_d \in \C$, we can use interpolation to write $h$ as a linear combination of $f_{\alpha_0}, f_{\alpha_1}, \ldots, f_{\alpha_d}$.
  Thus, $\DPD{h} \leq \sum_{0 \leq i \leq d} \DPD{f_{\alpha_i}} \leq (d+1) \cdot \DPD{f}$.
\end{proof}

\begin{definition}[Nisan Matrix~\cite{N91}]\label{defn:Nisan-matrix}
  For an $n$-variate polynomial $f(\vecx)$ of individual degree $d$, and a partition $S \sqcup T = [n]$, the \emph{$(S,T)$-Nisan matrix for $f$}, $M^{f}_{(S,T)}$, is a $(d+1)^{\abs{S}} \times (d+1)^{\abs{T}}$ matrix as follows.
  \begin{itemize}
      \item The rows are indexed by all the individual degree $d$ monomials over $\set{x_i \mid i \in S}$,
      \item The columns are indexed by all the individual degree $d$ monomials over $\set{x_j \mid j \in T}$,
      \item The entry $M^{f}_{(S,T)}[m,m']$ is the coefficient of the monomial $m \cdot m'$ in $f$.
  \end{itemize}
\end{definition}

\begin{theorem}[Nisan's characterization~\cite{N91}]\label{prop:Nisan-characterization}
  For any $n$-variate polynomial $f(\vecx)$, and any order $\sigma \in s_n$ on the variables, define $S_i = \set{\sigma(1),\ldots,\sigma(i)}$ and $T_i = \set{\sigma(i+1),\ldots,\sigma(n)}$ for each $i \in [n]$.\\
  Then the size of the smallest ROABP for $f$ in the order $\sigma$ is exactly $\sum_{i \in [n]} \mathrm{rank}(M^{f}_{(S_i,T_i)})$.
  Further, the width of the ROABP is exactly $\max_{i \in [n]} \mathrm{rank}(M^{f}_{(S_i,T_i)})$.
\end{theorem}

It is not difficult to see that for any polynomial, the Nisan matrix for any partition is a scaling of a sub-matrix of a matrix whose rows are all the partial derivatives of that polynomial.
This then leads to the following observation, which is a weaker and non-constructive version of \autoref{thm:main}.
\begin{observation}\label{obs:partials-and-Nisan}
    Let $n,d \in \N$ be arbitrary and $\F$ be any field of characteristic $0$ or greater than $d$.
    Then for any $n$-variate $f(\vecx)$ of individual degree $d$, and any partition $S \sqcup T = [n]$, $\mathrm{rank}(M^{f}_{(S,T)}) \leq \DPD{f}$.

    Thus, any polynomial $f$ has an ROABP in every order of width at most $\DPD{f}$.
\end{observation}

\begin{observation}\label{obs:pallindrome-order-dependence}
  The polynomial $(x_1+y_1)(x_2+y_2)\cdots(x_n+y_n)$ has width-$2$ ROABPs in the order $(x_1,y_1,\ldots,x_n,y_n)$, but requires width $2^n$ in the order $(x_1,\ldots,x_n,y_1,\ldots,y_n)$.
\end{observation}

\subsection{Concepts from algebra}

We will need a few concepts from elementary algebraic geometry; the reader may refer to any standard texts for more details on these concepts (e.g. \cite{CLO07, IK99}).

\begin{definition}[Ideal]\label{defn:ideal}
    For a set of polynomials $\set{f_1,\ldots,f_s} \subset \C[\vecx]$, the ideal generated by them is the smallest set of polynomials $I$ that satisfies the following.
    \begin{itemize}
        \item $\forall g \in \C[\vecx]$ and $\forall f \in I$, $fg \in I$.
        \item $\forall f, f' \in I$, we have $f + f' \in I$.
    \end{itemize}
    The ideal is denoted by $\inangle{\set{f_1,\ldots,f_s}}$.
\end{definition}

\begin{definition}[Variety of an ideal]\label{defn:variety-of-ideal}
    For an ideal $I \subset \C[\vecx]$, the \emph{variety of $I$}, written as $\V(I)$, is the largest set of points $V \subset \C^{\abs{\vecx}}$ such that $\forall f \in I$ and $\forall \veca \in V$, $f(\veca) = 0$.
\end{definition}

\subsubsection*{Derivative operators}

\begin{definition}[Derivative Operator]\label{defn:derivative-operator}
    A derivative operator is a linear combination of finitely many partial derivatives of the form $D = \sum_{i = 1}^r \alpha_i \partial_{m_i}$.
    It acts on polynomials naturally: $D f = \sum_{i = 1}^r \alpha_i \partial_{m_i} f$.

    Clearly, for any polynomial $g = \sum_m g_m m$, we can define a derivative operator $D_g = \sum_m g_m \partial_m$, and vice versa.
    Therefore, we always refer to a derivative operator as $D_g$ with an implicit polynomial $g$. 
\end{definition}

\begin{definition}[Closed space of derivative operators]\label{defn:closed-derivative-space}
    A vector space of operators $\Delta$ is said to be closed if for every $D_g \in \Delta$, and any monomial $m$ such that $g' := \partial_m g \not\equiv 0$, the corresponding operator $D_{g'}$ is also in $\Delta$.
\end{definition}

\begin{observation}\label{obs:props-derivative-operators}
  For any $f(\vect), g(\vect) \in \C[\vect]$, we have the following.
  \[ \inparen{D_f g}(\mathbf{0}) = \sum_{\vece} \coeff_{f}(\vect^{\vece}) \cdot \vece! \cdot \coeff_{g}(\vect^{\vece}) = \inparen{D_g f}(\mathbf{0})\]
\end{observation}

\paragraph*{Primary ideals and derivative operators}

The following result follows from the joint works of {\Moller}, Marinari and Mora~\cite{MMM93}, and {\Moller} and Stetter~\cite{MS95}.
A proof, with a statement as follows, can be found in \cite{RT22}. 

\begin{theorem}[{\cite{MMM93,MS95}}]\label{prop:MMM-correspondence}
  Let $J \subseteq \C[\vect]$ be an ideal with the variety $\V(J) = \set{\mathbf{0}}$, and suppose that the quotient ring $R_J := \sfrac{\C[\vect]}{J}$ is a $w$-dimensional vector space over $\C$.
  Then, there exists a $w$-dimensional $\C$-vector space of derivative of operators $\Delta(J)$ that characterizes the quotient ring $R_J$.

  That is, for any basis $\set{D_1,\ldots,D_w}$ of $\Delta(J)$, there is an \emph{invertible} matrix $M \in \C^{w \times w}$ such that for any polynomial $g(\vect) \in \C[\vect]$,
  \[
      \trans{[ D_1(g)(\mathbf{0}) \,\,\, D_2(g)(\mathbf{0}) \,\, \cdots \,\, D_w(g)(\mathbf{0}) ]} = M \cdot \cvector([g]),
  \]
  where $\cvector([g])$ is the coefficient vector of $[g]:=(g \bmod J)$.
\end{theorem}

The above correspondence works for any point $\veca \in \C^n$ in the variety; we will work with $\mathbf{0}$ to keep the exposition simple, since that is the only case relevant for our application.

\begin{definition}[Operator space of an ideal]\label{defn:operator-space-of-ideal}
  For an ideal $J \subseteq \C[\vect]$ with $\V(J) = \set{\mathbf{0}}$, we define the corresponding space of derivative operators $\Delta(J)$ as follows.
  \[
      \Delta(J) := \set{ D_g \in \C[\partial\vect] : \forall h \in J, (D_g h)(\mathbf{0}) = 0 } \qedhere
  \]
\end{definition}

\begin{definition}[Ideal of an operator space]\label{defn:ideal-of-operator-space}
  Let $\Delta$ be a closed space of derivative operators.
  We define the corresponding annihilating ideal (at the point $\mathbf{0}$), denoted by $\I_{\mathbf{0}}(\Delta)$ as follows.
  \[
      \I_{\mathbf{0}}(\Delta) := \set{ h \in \C[\vect] : \forall D \in \Delta, (D h)(\mathbf{0}) = 0 } \qedhere
  \]
\end{definition}

\subsection{Multiplication tables: from ideals to matrices}

\subsubsection*{Univariate ideals}
Let $J = \inangle{p(t)} \subseteq \C[t]$ be an ideal, and consider the quotient ring $R := \sfrac{\C[t]}{J}$.
If $p(t)$ has degree $d$, then the \emph{multiplication table} for $t$ in the ring $R$, is a $d \times d$ matrix whose \emph{minimal polynomial} is $p(t)$.
Such a matrix, say $A$, can easily be defined by setting $A_{i,j} = \coeff_{t^j}(\insquare{t \cdot t^i})$, for all $0 \leq i,j \leq (d-1)$.
Here, $[t \cdot t^i]$ is $(t^{i+1} \bmod J)$. 

For instance, when $p(t) = t^5 - 10 t^4 - 7 t^3 + 2 t^2 - 3$, the multiplication table would be the following $5 \times 5$ matrix; it can be checked that $p(t)$ is indeed the minimal polynomial of $A$.
\[
    A = 
    \begin{bmatrix}
        0 & 1 & 0 & 0 & 0 \\
        0 & 0 & 1 & 0 & 0 \\
        0 & 0 & 0 & 1 & 0 \\
        0 & 0 & 0 & 0 & 1 \\
        3 & 0 & -2 & 7 & 10 \\
    \end{bmatrix}
\]

Further, $A$ satisfies that $g(A)_{i,j}$ is exactly $\coeff_{t^j}(\insquare{g(t) \cdot t^i})$, for any $g(t) \in \C[t]$.
In particular, this means that the first row of $g(A)$ is precisely the coefficient vector of $\insquare{g(t)}$.

\subsubsection*{Multivariate ideals}

One key change when we move to the multivariate setting, is that there is no inherent ordering on the monomials; so we have to choose one.
A \emph{monomial ordering} is any `total order' on monomials, which respects divisions and has $1$ as the least monomial.
We will work with the degree-wise lexicographical ordering (``deg-lex'') with respect to $t_1 \prec t_2 \prec \cdots \prec t_r$.
We use this monomial ordering to uniquely identify the leading (``greatest'') and trailing (``least'') monomials in any polynomial in $\C[t_1,\ldots,t_r]$.

This then allows us to identify a set of leading monomials of the ideal, and then define what is called a normal set and the quotient ring corresponding to the ideal, as follows.
\begin{definition}[Leading monomials and normal set]\label{defn:leading-monomials-normal-set}
    Given an ideal $I \subset \C[\vecx]$, and a monomial ordering, the set of its leading monomials is defined as $\LM(I) := \set{ \LM(f) \mid f \in I }$.

    The complement of $\LM(I)$ is called the \emph{normal set of $I$}, denoted by $\normalset(I)$.
\end{definition}

\begin{definition}[Quotient ring]\label{defn:quotient-ring}
    For any ideal $I \subset \C[\vecx]$, and any polynomial $g \in \C[\vecx]$, we can define $g \bmod I$ to be the polynomial $g_0$ all of whose monomials are from $\normalset(I)$ and which satisfies $g - g_0 \in I$.
    We denote the polynomial $g \bmod I$ by $[g]$ when the ideal is clear from the context.
    
    We can thus define the \emph{quotient ring} corresponding to $I$ denoted by $\sfrac{\C[\vecx]}{I}$, by reducing each polynomial in $\C[\vecx]$ modulo $I$. 
\end{definition}

Intuitively, the quotient ring $\sfrac{\C[\vect]}{J}$ is obtained by ``setting all polynomials in $J$ to zero''.
Then any monomial from $\LM(J)$ can be written in terms of those in the normal set, and the polynomials in the quotient ring are supported entirely on the monomials from $\mathrm{N}(J)$.

Let the normal set of $J \in \C[t_1,\ldots,t_r]$ be $\set{m_1,\ldots,m_w}$, where $1 = m_1 \prec m_2 \prec \cdots \prec m_w$.
The multiplication tables $A_1,A_2,\ldots,A_r$ for $J$ are then the $w \times w$ matrices that satisfy the following, for every $\ell \in [r]$, and all $i, j \in [w]$.
\[
  A_{\ell}(i,j) = \coeff_{m_j}(\insquare{t_{\ell} \cdot m_i})
\]
Just as before, for any polynomial $g(\vect)$, we have that $g(A_1,\ldots,A_r)(i,j) = \coeff_{m_j}(\insquare{g \cdot m_i})$.
Thus, the first row of any matrix of the form $g(A_1,\ldots,A_r)$ is just the coefficient vector of $g(\vect) \bmod J$, for any polynomial $g$.

\section{Constructing commutative ROABPs from apolarities}

\subsection{Apolar ideal of a polynomial}

\begin{definition}[Apolar Ideal]\label{defn:apolar-ideal}
  Let $f(t_1,\ldots,t_n)$ be a homogeneous polynomial of degree $d$.
  The \emph{apolar ideal of $f$} is defined as follows.
  \[
      f^{\perp} := \inangle{ \set{ h(\vect) \in \C[\vect]: D_h f \equiv 0} } \qedhere
  \]
\end{definition}

\begin{observation}\label{obs:variety-of-apolar-ideal}
  For any polynomial $f(\vect)$, the variety of its apolar ideal, $\V(f^{\perp})$, is a single point $\mathbf{0}$.
\end{observation}
\begin{proof}
  For each $i \in [n]$, $t_i^{d + 1} \in f^{\perp}$ where $d = \deg(f)$; so the variety is contained in $\set{\mathbf{0}}$.
  Also, when $f \not\equiv 0$, any polynomial in $f^{\perp}$ has a zero constant term because a nonzero polynomial cannot be linearly dependent on its derivatives. 
  Hence, $\V(f^{\perp}) = \set{\mathbf{0}}$. 
\end{proof}

Note that the apolar ideal is a polynomial ideal, but is defined using derivative operators.
The apolar ideal of $f$ is related to its partial derivatives in the following way.

\begin{lemma}[Apolar ideal and partial derivatives]\label{lem:apolarity-and-partials}
  Let $f(\vect)$ be a homogeneous polynomial with the set $\set{g_1,g_2,\ldots,g_w}$ being a basis for its space of partial derivatives of all degrees.
  For the corresponding closed space of derivative operators $\Delta_f := \mathrm{span}_{\C} \set{D_{g_1},D_{g_2},\ldots,D_{g_w}}$, and its annihilating ideal $J := \I_{\mathbf{0}}(\Delta_f)$, we have that $J = f^{\perp}$ and equivalently, $\Delta_f = \Delta(f^{\perp})$.
\end{lemma}
\begin{proof}
  We can assume that $f = g_1$ without loss of any generality.
  
  \begin{description}\itemsep10pt
    \item[$f^{\perp} \subseteq J$.]
                Let $h \in f^{\perp}$, and say $g = \partial_m f$ is some arbitrary partial derivative of $f$, so $D_g \in \Delta_f$.\\
                Now $\inparen{D_{g} h}(\mathbf{0}) = \sum_{\vece} \coeff_h(\vect^{\vece}) \cdot \vece! \cdot \coeff_{g}(\vect^{\vece}) = \inparen{D_{h} g}(\mathbf{0})$, from \autoref{obs:props-derivative-operators}.
                Further, $D_{h} g = D_h \partial_{m}f = D_{(h \cdot m)} f$, and since $m \cdot h \in f^{\perp}$, $D_g h = D_{(h \cdot m)} f = 0$.
                Since our choice of $g$ and $h$ was arbitrary, this is true for each $g \in \Delta_f$, and each $h \in J$, showing that $f^{\perp} \subseteq J$.
    \item[$J \subseteq f^{\perp}$.]
                Let $h \in J$ be arbitrary, and consider $D_h f = \sum_{\vece} \sum_{\vece'} \coeff_{h}(\vect^{\vece}) \cdot \coeff_{f}(\vect^{\vece'}) \cdot \partial_{\vece}(\vect^{\vece'})$.
                \begin{align*}
                  D_h f &= \sum_{\vece} \sum_{\vece' \geq \vece} \coeff_{h}(\vect^{\vece}) \cdot \coeff_{f}(\vect^{\vece'}) \cdot \partial_{\vece}(\vect^{\vece'})\\
                        &= \sum_{\vece} \sum_{\vece' \geq \vece} \coeff_{h}(\vect^{\vece}) \cdot \coeff_{f}(\vect^{\vece'}) \cdot \frac{\vece'!}{(\vece' - \vece)!} \cdot (\vect^{\vece' - \vece})\\
                        &= \sum_{\vece_0 := \vece' - \vece}  \insquare{\sum_{\vece} \coeff_{h}(\vect^{\vece}) \cdot (\vece + \vece_0)! \cdot \coeff_{f}(\vect^{\vece + \vece_0})\cdot \inparen{\frac{\vect^{\vece_0}}{\vece_0!}}}
                \end{align*}
                Now let $g_0 := \partial_{\vece_0}(f)$, and note that $\coeff_{g_0}(\vect^{\vece}) = \frac{(\vece_0 + \vece)!}{\vece!} \cdot \coeff_{f}(\vect^{\vece_0 + \vece})$.
                Therefore, we can further simplify our expression for $D_h f$ as follows.
                \begin{align*}
                  D_h f &= \sum_{\vece_0} \frac{\vect^{\vece_0}}{\vece_0!} \cdot \inparen{\sum_{\vece} \coeff_{h}(\vect^{\vece}) \cdot \vece! \cdot \coeff_{g_0}(\vect^{\vece})} & \\
                        &= \sum_{\vece_0} \frac{\vect^{\vece_0}}{\vece_0!} \cdot \inparen{D_{g_0} h}(\mathbf{0}) & \text{(Using \autoref{obs:props-derivative-operators})}\\
                        &= \sum_{\vece_0} 0 \cdot \frac{\vect^{\vece_0}}{\vece_0!} \equiv 0 & \text{($D_{g_0} \in \Delta_f$ and $h \in J$)}
                \end{align*}
                Thus, $h \in f^{\perp}$. \qedhere
  \end{description}
\end{proof}

\subsection{Proof of Theorem \ref{thm:main}}

We are now ready state the general recipe for constructing a commutative ROABP for any homogeneous $f(x_1,\ldots,x_n)$.
We start by defining the following polynomial over $\vecx$, and an auxiliary set of variables $\vect = \set{t_1,\ldots,t_n}$, which is the product of the degree-$d$-truncations of the Taylor series of $e^{t_i x_i}$s.
\begin{equation}\label{eq:base-polynomial}
  G(\vecx,\vect) := \prod_{i = 1}^{n} \inparen{1 + t_i x_i + \frac{1}{2!} t_i^2 x_i^2 + \cdots  + \frac{1}{(d-1)!} t_i^{d-1} x_i^{d-1} + \frac{1}{d!} t_i^d x_i^d}
\end{equation}

\begin{observation}\label{obs:base-polynomial-works}
  Let $D := f(\partial t_1,\partial t_2,\ldots,\partial t_n)$. Then $(D \circ G) = f(\vecx)$.
\end{observation}

\begin{lemma}\label{lem:ROABP-for-f}
  Suppose $f(\vecx)$ is a homogeneous polynomial with dimension of partial derivatives exactly $w$.
  Then, there exists a vector $\vecv \in \C^w$, such that for the multiplication tables $A_1,\ldots,A_n$ of the apolar ideal $f^{\perp}$, we have that
  \[
      \sum_{j \in [w]} v_j \cdot G(A_1,\ldots,A_n,x_1,\ldots,x_n)[1,j] = f(\vecx). \qedhere
  \]
\end{lemma}
\begin{proof}
  Since $A_1,\ldots,A_n$ are multiplication tables of $f^{\perp}$, we get that the first row of $G(\vecA,\vecx)$ is exactly the coefficient vector of $(G(\vect,\vecx) \bmod f^{\perp}(\vect))$ which is an object in $ \sfrac{\C[\vect][\vecx]}{f^{\perp}(\vect)} = (\sfrac{\C[\vect]}{f^{\perp}})[\vecx]$.
  
  Next, suppose that $\set{g_1(\vect),\ldots,g_w(\vect)}$ is a basis for the partial derivatives of $f(\vect)$.
  Then by \autoref{lem:apolarity-and-partials}, we know that $\Delta(f^{\perp})$ has a basis given by the operators $\set{D_{g_1},\ldots,D_{g_w}}$.
  Also, the variety of $f^{\perp}$ is exactly the singleton set $\set{\mathbf{0}}$.
  Thus, by \autoref{prop:MMM-correspondence}, the coefficients of $\insquare{G(\vect,\vecx)} := G(\vect,\vecx) \bmod f^{\perp}(\vect)$ are spanned by $D_{g_1}(G)(\mathbf{0}), D_{g_2}(G)(\mathbf{0}), \ldots, D_{g_w}(G)(\mathbf{0})$.
  More importantly, the set of $D_{g_i}(G)$'s is spanned by the coefficients of $\insquare{G(\vect,\vecx)}$, and further, $D_{g_1}(G) = D_f(G) = f(\vecx)$.

  Thus, the vector $\vecv$ can be obtained from the matrix $M$ guaranteed by \autoref{prop:MMM-correspondence}, as claimed.
\end{proof}

This proves the main theorem, restated below.

\MainTheorem*

\begin{remark}\label{rmk:connection-to-duality}
    Note that Saxena's proof of the duality trick~\cite[Lemma 1]{S08b} can be seen as starting with the same ``template polynomial'' as \eqref{eq:base-polynomial}, homogenizing it through interpolation, and then just evaluating it on the ``points'' given by the Waring decomposition for $f$.
    Since it is known by the \emph{Apolarity lemma} (see e.g. \cite{IK99}) that the points given by a Waring decomposition of $f$ define a radical ideal $J$ that sits inside $f^{\perp}$, the action of evaluating on those points can be viewed as going modulo $J$.
    
    In that sense, our proof generalizes this method by directly going modulo $f^{\perp}$.
    As this uses a less strict property of $f$, the resulting expression is less simple, and is therefore a commRO instead of a diagRO.
\end{remark}

\section{The Determinant}\label{sec:commRO-for-determinant}

In this section, we will use the proof of \autoref{thm:main} to construct explicit commutative ROABPs. In particular, we will construct a commutative ROABP for the determinant (Det$_n$) of width $2^{\Theta(n)}$.

The choice of this example is deliberate, as the determinant is the only candidate where there is an asymptotic gap between Waring rank upper bounds  and  partial derivative dimension. 

The determinant of $n$-dimensional symbolic matrix has partial derivative dimension exactly $\binom{2n}{n}=2^{\Theta(n)}$. But, the best upper bound for the Waring rank of the determinant is $2^{O(n \log n)}$.
In fact, there are reasons to believe that the Waring rank of the determinant is $2^{\omega(n)}$.
This is due to the fact that the set-multilinear depth-3 complexity or \emph{Tensor rank} of $\Det_n$, and (to the best of our knowledge) even the best constant-depth multilinear formula that we know of for the determinant, is $2^{O(n \log n)}$.
See \cite{KM18, R10} for details on tensor rank and syntactic multilinear formulas of the determinant.

For $\Det_n$, \autoref{thm:main} directly gives a commutative ROABP of width $2^{\Theta(n)}$. 
We show the explicit calculations behind this commutative ROABP below.
Let's recall what our overall step-by-step process will be for any polynomial $f \in \C[\vecx]$: 
\begin{enumerate}
  	\item Compute the closed derivative space $\Delta= \partial^{< \infty} f$, the apolar ideal corresponding to it $I_{\mathbf{0}}(\Delta)= f^{\perp}$ and the normal set of $\mathrm{N}({f^{\perp}})$. Let, $m:=\abs{\mathrm{N}({f^{\perp}})}=\abs{\Delta}$.
    \item Compute the multiplication tables ($M_i$) corresponding to each of the variables. 
    \item The final commutative ROABP of    $f \equiv \trans{\veca} \cdot \prod_{i \in [n] } (1 + x_{i} M_{i}) \cdot \vecb$ for $\veca, \vecb \in \C^m$.
\end{enumerate}

\subsection{Derivative Space, Apolar Ideal, and its Normal Set}
In this subsection, we will state and discuss some facts about the derivative space, apolar ideal, and the normal set of the apolar ideal of the determinant.
We will use $X$ to denote the $n \times n$ symbolic matrix, that is, $X = (x_{i,j})_{i,j \in [n]}$. Similarly, let $U = (t_{i,j})_{i,j \in [n]}$ be a symbolic matrix in $t$-variables. Let $S, T$ be arbitrary subsets of $[n]$. We will denote the minor (of $X$) picked by selecting rows from $S$ and columns from $T$ by $X_{S,T}$.

We will start with the well-known fact that the derivative space of determinant is just the determinant of its minors. Formally, the following set is a basis of $\partial^{< \infty} \Det_n(X), $ \[  
\left\{ \Det (X_{S,T}) : \text{ for } S, T \subseteq [n] \text{ such that } \abs{S} = \abs{T} \right\}. \]

The apolar ideal of the determinant is generated by permanents of \( 2 \times 2 \) minors  and certain unacceptable degree two monomials, as stated formally below.

\begin{theorem}[{e.g. \cite[Theorem 2.12]{S15b}}]
\(\Det_n^{\perp}(X) = \langle \mathcal{P}_X, \mathcal{U}_X \rangle\), where \(\mathcal{P}_X\) is the collection of permanents of all \(2 \times 2\) minors of \(X\), and \(\mathcal{U}_X\) denotes all quadratic unacceptable monomials, that is,  monomials that don't divide any monomial in the support of \(\Det_n(X)\).
\end{theorem}

Now, for the normal set computation, we will focus on the degree-wise lexicographical ordering of monomials (``deg-lex''); the ordering on variables is in the ``row-major'' form: \[x_{1,1} \succ x_{1,2} \succ \ldots \succ x_{1,n} \succ x_{2,1} \succ \ldots \succ x_{n,n}.\]
The trailing monomial of $\Det(X_{S,T})$ in this ordering is just the product of the anti-diagonal entries of $X_{S,T}$. To see this, note that the only variable you can pick from the first row is the last element. Now that we have picked something from the last column and the first row, we can strip them off as none of the variables can contribute anymore. Now focus on the resulting minor (after stripping) and proceed by induction. Let's denote this trailing monomial by $\tau_{S,T} := \text{anti-diag}(X_{S,T})$.

We now claim that the normal set of $J$ is just a collection of these anti-diagonal monomials corresponding to all minors, as follows.
\[ 
\mathrm{N}(J)=\{ \text{anti-diag}(X_{S,T}) : \text{ for } S, T \subseteq [n] \text{ such that } \abs{S} = \abs{T} \} 
\]
To see this, observe that any non-anti-diagonal monomial (for any minor) will be a multiple of the leading term of a $2 \times 2$ minor's permanent and thus will be in $\LM(\Det^{\perp})$. At the same time, the anti-diagonal monomial will never be a multiple of such terms, so it is never in $\LM(\Det^{\perp})$. Note that here, $\abs{\Delta} = \abs{\mathrm{N}(J)} = \binom{2n}{n}$.

\subsection{Multiplication tables for the apolar ideal}

Let $A_{i,j}$ be the matrix corresponding to $t_{i,j}$ with dimension $\abs{\mathrm{N}(J)} \times \abs{\mathrm{N}(J)}$.    For any row of $A_{i,j}$, indexed by $(S,T)$ such that $\abs{S}=\abs{T}$, we have that 
\[
 row(S, T)(A_{i,j}) = \begin{cases} 
0 & \text{if } i \in S \text{ or } j \in T \\
\operatorname{sgn}(\tau_{S', T'}) \cdot {\operatorname{sgn}(\tau_{S, T} \cdot x_{i,j})} \cdot \tau_{(S \cup \{i\}, T \cup \{j\})} & \text{if } i \notin S, j \notin T 
\end{cases}.
\]
Here, $ \operatorname{sgn}$ of any monomial denotes the sgn of its corresponding permutation (obtained by viewing $S,T \equiv \set{1,\ldots,\abs{S}}$).
To see this, note that by definition $ row(S, T)(A_{i,j})$ is just the coefficient vector of $(t_{i,j} \cdot \tau_{S,T} \bmod J)$. 

Thus, as discussed (in proof of \autoref{thm:main}) we get 
\[\Det_n(X)= \trans{\veca} \cdot \prod_{i,j \in [n]}(I + A_{i,j}x_{i,j}) \cdot \vecb \qquad \text{for $\veca, \vecb \in \C^{\binom{2n}{n}}$.}\] 

Below, we give an explicit description of the commutative ROABP for $\Det_2$.
Here, the normal set is $\mathrm{N}(\Det_2^{\perp})=\left\{ 1, x_{1,1}, x_{1,2}, x_{2,1}, x_{2,2}, x_{1,2} x_{2,1} \right\}$.
Running our analysis to compute the multiplication tables followed by then replacing it in the template polynomial yields that the $\Det_2(X)$ is the $(1,n)$-th entry of the product of the following four matrices (in any order).
\[
M_{1,1} =
\begin{pmatrix}
1 & x_{1,1} & 0 & 0 & 0 & 0 \\
0 & 1 & 0 & 0 & 0 & 0 \\
0 & 0 & 1 & 0 & 0 & 0 \\
0 & 0 & 0 & 1 & 0 & 0 \\
0 & 0 & 0 & 0 & 1 & -x_{1,1} \\
0 & 0 & 0 & 0 & 0 & 1 \\
\end{pmatrix}
,\,\, M_{1,2} =
\begin{pmatrix}
1 & 0 & x_{1,2} & 0 & 0 & 0 \\
0 & 1 & 0 & 0 & 0 & 0 \\
0 & 0 & 1 & 0 & 0 & 0 \\
0 & 0 & 0 & 1 & 0 & x_{1,2} \\
0 & 0 & 0 & 0 & 1 & 0 \\
0 & 0 & 0 & 0 & 0 & 1 \\
\end{pmatrix},
\]
\[
M_{2,1} =
\begin{pmatrix}
1 & 0 & 0 & x_{2,1} & 0 & 0 \\
0 & 1 & 0 & 0 & 0 & 0 \\
0 & 0 & 1 & 0 & 0 & x_{2,1} \\
0 & 0 & 0 & 1 & 0 & 0 \\
0 & 0 & 0 & 0 & 1 & 0 \\
0 & 0 & 0 & 0 & 0 & 1 \\
\end{pmatrix}
,\,\, M_{2,2} =
\begin{pmatrix}
1 & 0 & 0 & 0 &  x_{2,2} & 0 \\
0 & 1 & 0 & 0 & 0 & -x_{2,2} \\
0 & 0 & 1 & 0 & 0 & 0 \\
0 & 0 & 0 & 1 & 0 & 0 \\
0 & 0 & 0 & 0 & 1 & 0 \\
0 & 0 & 0 & 0 & 0 & 1 \\
\end{pmatrix}.
\]

\subsection{Commutative Set-multilinear ABP}\label{subsec:Comm SML ABP}
The  commutative matrices $A_{i,j}$  generated in the previous subsections using multiplication tables from $\Det_n^{\perp}$ can in fact be used to  design a  \textit{commutative} set-multilinear ABP for $\Det$ as well. 

The benefit of studying this model stems from the fact that if we could somehow ``diagonalize'' these commutative matrices, then that would indeed give a set-multilinear depth-3 representation of the determinant of size \(2^{O(n)}\). That, in turn, would yield that the Waring rank of the determinant is also \(2^{O(n)}\). By ``diagonalizable,'' we simply mean if the above matrices can be replaced by diagonal matrices with at most a polynomial blow-up in the dimension.

In fact, the commutative matrices constructed for $\Det_n$ are provably not diagonalizable by invertible transformations. But for $\Perm_n$, the matrices that we will get by running our entire analysis  are indeed ``diagonalizable''! In the sense that we can replace them by diagonal matrices of similar dimension to compute $\Perm_n$.

Let $ \sqcup _{j \in [d]} \vecx_j$ be a partition of the set $\vecx$ of input variables. Then a polynomial is set-multilinear under partition $\sqcup_{j \in [d]} \vecx_j$ if  each monomial of the polynomial picks up \textit{exactly} one variable from each part in the partition. Note that, $\Det$ is set-multilinear w.r.t. the variable partition being the row variables (or column variables).

\begin{definition}[Set-multilinear ABP (smABP)]\label{defn:smABP}
  Let $n,d,w \in \N$, and let $f(\vecx)$ be an $n$-variate set-multilinear polynomial under the partition $\vecx_1 \sqcup \vecx_2 \sqcup \cdots \sqcup \vecx_d$. 
  We say that $f$ has a width $w$ set-multilinear ABP\footnote{Strictly speaking this defines \textit{ordered} set-multilinear algebraic branching programs, but we drop this detail for brevity.}, if there exists a permutation $\sigma \in s_d$ for which there exist matrices $\set{A_{j,k}}$ in $\C^{w \times w}$ for all $j \in [d]$ and $1 \leq k \leq \abs{\vecx_j}$, and vectors $\vecu, \vecv \in \C^w$, such that the following holds.
  \begin{align*}
      f(\vecx) &= \trans{\vecu} \cdot M_{\sigma(1)}(\vecx_{\sigma(1)}) \cdot M_{\sigma(2)}(\vecx_{\sigma(2)}) \cdots M_{\sigma(d)}(\vecx_{\sigma(d)}) \cdot \vecv,\\
      &\text{where for all $j \in [d]$,}\\
      M_{j}(\vecx_{j}) &=   A_{j,1} x_{j,1} + A_{j,2} x_{j,2} + \cdots + A_{j,\abs{\vecx_j}} x_{j,\abs{\vecx_j}}.
  \end{align*}
  We call the matrices $\set{A_{j,k}}$ the \emph{coefficient matrices} of the smABP.
\end{definition}

\begin{definition}[Commutative smABP]\label{defn:comm-smABP}
  An smABP is said to be a \emph{commutative smABP} if all its coefficient matrices pairwise commute with each other.
\end{definition}


Now, we will show that by simply changing the template polynomial (from \eqref{eq:base-polynomial}) appropriately, we can get a commutative set-multilinear ABP representation for $\Det_n$.
\begin{equation}\label{eq:base-polynomial-sml}
  \text{Define, } G(\vecx,\vect) := \prod_{i = 1}^{n} \left( \sum_{j \in [n]}t_{i,j} x_{i,j}\right).
\end{equation}

Again, just like \autoref{obs:base-polynomial-works}, we have that $\Det_n(\partial_{t_{1,1}}, \partial_{t_{1,2}}, \ldots, \partial_{t_{n,n}}) \circ G = \Det_n(X)$.
And this gives that,
\[ \Det_n(X) = \trans{\veca} \cdot \prod_{i = 1}^{n} \left( \sum_{j \in [n]}A_{i,j} x_{i,j} \right) \cdot \vecb \qquad \text{for some $\veca, \vecb \in \mathbb{C}^{\binom{2n}{n}}$.}\]

We remark that the above analysis works for any set-multilinear polynomial, along the same lines as the proof of \autoref{thm:main}. This directly gives us the following theorem.

\commSMABP*

\section{Discussion}

In summary, we utilize the knowledge of commuting matrices outlined in \cite{RT22} to provide a generic recipe for explicit constructions of commutative branching programs.
For the specific setting of ROABPs, this improves upon the earlier known connection (\autoref{obs:partials-and-Nisan}) and takes us a step closer towards answering \autoref{quest:DPD-captures-Waring}.
An immediate direction for further study is the following.

Can we show any bounds on the commRO width of a polynomial in terms of its diagRO width?
In addition to shedding further light on \autoref{quest:DPD-captures-Waring}, such a result should also provide us with a new hardness measure for structured ROABPs and possibly even depth 3 powering circuits, particularly if the bound is a super-linear lower bound on diagRO width.
As alluded to in the introduction, perhaps a new hardness measure against diagRO or depth 3 powering circuits is what is required to fully derandomize blackbox PIT for the model.
A concrete way in which this is true is that a polynomial time blackbox PIT for width-$w$, degree-$d$, $O(\log dw)$-variate diagRO would give a polynomial time blackbox PIT for depth 3 powering circuits~\cite[Lemma 2.12]{BS21}.
This gives the lower bound question a much larger and more interesting context.

\section*{Acknowledgements}

VB thanks Rafael Oliveira and Abhiroop Sanyal for numerous insightful discussions.
AT thanks Prerona Chatterjee, C Ramya, and Ramprasad Saptharishi for several fruitful discussions about ROABPs over the recent years.
AT is also deeply grateful to Ramprasad Saptharishi, Susmita Biswas and Lulu, for hosting him during a part of this work.

\bibliographystyle{customurlbst/alphaurlpp}
\bibliography{masterbib/references,masterbib/crossref}

\appendix

\end{document}